\documentclass[12pt]{iopart}

\usepackage{iopams}
\usepackage{amssymb}
\usepackage{graphicx}
\usepackage{float}
\newcommand{\ket}[1]{\left| #1 \right\rangle}
\begin{document}

\title{State-dependent, addressable subwavelength lattices with cold atoms}

\author{W Yi, A J Daley, G Pupillo and P Zoller}
\address{Institute for Quantum Optics and Quantum Information of the Austrian
Academy of Sciences, A-6020 Innsbruck, Austria} \address{Institute for
Theoretical Physics, University of Innsbruck, A-6020 Innsbruck, Austria}
\ead{wei.yi@uibk.ac.at}

\begin{abstract}
We discuss how adiabatic potentials can be used to create addressable lattices
on a subwavelength scale, which can be used as a tool for local operations and
readout within a lattice substructure, while taking advantage of the faster
timescales and higher energy and temperature scales determined by the shorter
lattice spacing. For alkaline-earth-like atoms with non-zero nuclear spin,
these potentials can be made state dependent, for which we give specific
examples with $^{171}$Yb atoms. We discuss in detail the limitations in
generating the lattice potentials, in particular non-adiabatic losses, and show
that the loss rates can always be made exponentially small by increasing the
laser power.

\end{abstract}
\pacs{32.80.Qk, 03.67.Lx}
\maketitle

\section{Introduction}
Optical lattices in 3D generated by counterpropagating laser fields provide a
periodic array of microtraps for cold atoms, and implement quantum lattice
models governed by Hubbard Hamiltonians \cite{opticallattices1}. This is of
interest both in the study of strongly correlated systems from condensed matter
physics, and in the implementation of quantum information processing. Two key
elements of cold atom dynamics in 3D optical lattices are: (i) The lattice
spacing is bounded below by the wavelength $a\equiv\lambda/2$. Therefore, all
the relevant time, energy and temperature scales are determined by the recoil
energy, $E_R=\hbar^2 k^2 / (2m)$, where $k=2\pi/\lambda$ is the wavenumber, and
$m$ is the the mass of the atom. In particular, the hopping matrix element
between neighbouring sites within the region of validity of the single-band
Hubbard model is limited by $J\ll E_R$. This limits all corresponding
timescales, for example the timescale over which entanglement can be generated
using exchange interactions \cite{Foelling,exchangegate}. (ii) Varying the
laser parameters provides \emph{global} control of the lattice potential, but
development of tools for local operations and readout remains a major challenge
in light of applications in quantum computing and quantum simulation
\cite{briegelreview}. Here we show how adiabatic potentials can be used to
address these issues by creating addressable, state-dependent sub-wavelength
lattices. Each lattice period can be divided into a set of wells, which can be
frequency-addressed via their offset energies, reminiscent of quantum computing
schemes with electrons in arrays of electrically gated quantum dots
\cite{Loss}.

In recent seminal experiments \cite{Foelling,exchangegate}, optical
superlattices have been used to provide addressability and to entangle atoms in
a series of double-wells in one parallel operation. Our goal here is to achieve
similar control on sub-wavelength scales by coupling optical lattice potentials
with dressing fields, with the benefit that all corresponding timescales are
significantly faster, and energy and temperature scales significantly higher.
Single particle effects of such dressed potentials \cite{deutsch98,
raithelprl,deutsch00,Goerlitz01,Dubetsky02,Ritt06,Zhang06,Ravaine06,add2} have
been studied, e.g., in the context of lithography \cite{Pfau}, laser cooling
\cite{Kaza,GrimmJETP}, and Radio Frequency (RF)-dressed potentials on atom
chips \cite{hoff06}. Here we show how the adiabatic potentials, created by
coupling optical lattice potentials with additional dressing fields, can be
used for subwavelength control in the context of manybody physics. We will
demonstrate how adiabatic potentials can be used to generate state-dependent,
addressable subwavelength lattices. In such a subwavelength lattice, each
lattice period is subdivided into several potential wells, for which the
relative well depth can be tuned by either applying external magnetic or
electric fields, or by adjusting the laser parameters, e.g. the Rabi-frequency
and detuning.  Atoms loaded into different wells within the subwavelength
structure  can also be frequency-addressed via their different offset energies.

In view of recent seminal progress with alkaline-earth-like atoms, e.g. in the
context of developing optical clocks \cite{zeemanshift1,zeemanshift2}, in
many-body cold atom physics \cite{takasu07}, and in applications in quantum
information \cite{deutsch07,babcock}, we investigate, in particular, the
possibility for generating sub-wavelength potentials with these species. The
existence of metastable triplet states makes possible the production of
near-resonant optical lattices, which allow for the creation of state-dependent
sub-wavelength lattices. We show how this can be used to generate and
manipulate entanglement, making use of the spin dependence and/or shorter
tunneling times within the subwavelength structure. Of course, in the context
of adiabatic potentials, the question of non-adiabatic losses will always
arise. We analyse these losses in detail in the present context, and conclude
that they can always be suppressed at the expense of using additional laser
power.

\begin{figure*}[tbp]
\includegraphics[width=15cm]{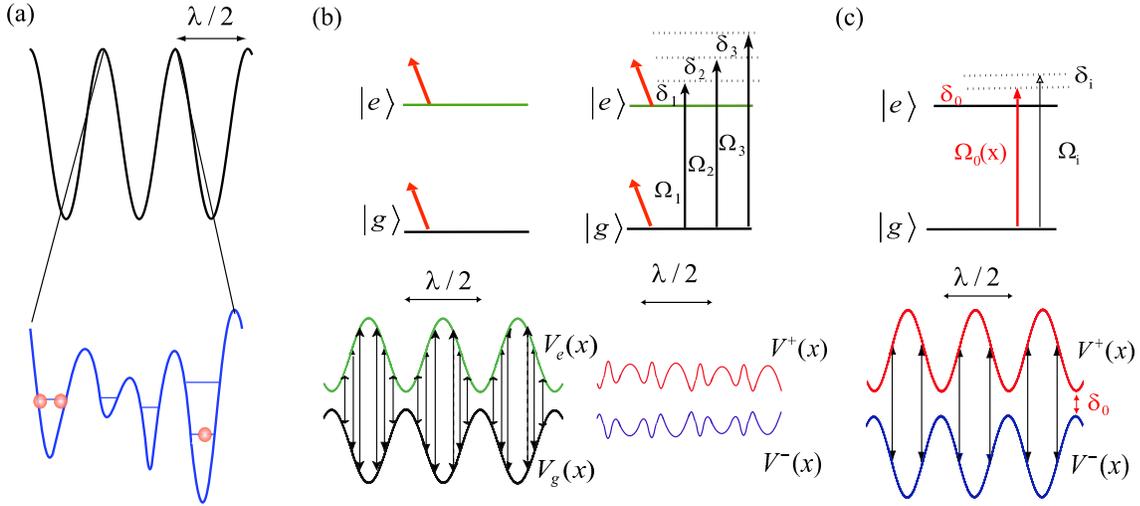}
\caption[Fig. 1]{Sub-wavelength lattice with off-resonant and near-resonant
$\lambda/2$ background potentials (see text). (a) Schematic of a controllable
subwavelength lattice, where each lattice period of an initial $\lambda/2$ is
subdivided into several potential wells whose position and well depth can be
controlled by changing the laser parameters. Note that addressability in the
subwavelength lattice is modulo $\lambda/2$, with $\lambda$ the wavelength of
the laser generating the background lattice potentials. (b) (left) Level
structure and laser configurations for generating background potentials
$V_{g,e}(x)$ in the off-resonant scheme. (right) Dressing fields with
Rabi-frequency $\Omega_i$($i=1,2,3,...$) and detuning $\delta_i$ are applied to
couple the background potentials to generate subwavelength lattice
$V^{\pm}(x)$. The separation between neighbouring wells is approximately
$\lambda/2(n+1)$ with $n$ dressing fields. (c) Level structure and laser
configurations for generating near-resonant subwavelength potentials in a two
level system. The counteroscillating background potentials $V^{\pm}(x)$ are now
induced by directly coupling the two levels $|g\rangle$ and $|e\rangle$ with
standing wave fields with Rabi-frequency $\Omega_0(x)$ and detuning
$\delta_0$(red). Dressing fields with Rabi-frequency $\Omega_i$($i=1,2,...$)
couple the background potentials to generate the subwavelength potential.}
\end{figure*}

This paper is organised as follows: We first give a general description in
Sec.~\ref{sec2} of two schemes to create adiabatic potentials with two-level
atoms. In Sec.~\ref{sec3}, we extend the idea to multi-level atoms, describing
the additional control that is present when these lattices are generated for
real atoms. In particular we discuss possibilities for state-dependent
potentials with alkaline-earth-like atoms. In Sec.~\ref{sec4}, we add
interactions between atoms and give examples for possible applications of
subwavelength lattices in quantum simulation and quantum information. This
includes a discussion of how entanglement within the subwavelength structure
can be generated and controlled. In Sec.~\ref{sec5}, we analyse the
Landau-Zener-type loss rate from the adiabatic potentials in detail and discuss
how it should be suppressed. In appendix A we give additional details on the
clock transition in alkaline-earth-like atoms, which are essential for the
generation of subwavelength lattices with these atoms.

\section{Generating subwavelength lattice potentials\label{sec2}}

In this section we introduce the basic ideas behind addressable subwavelength
potentials for two-level atoms. This will be generalised to multi-level atoms
in Sec.~\ref{sec3}, where we also discuss their implementation with
alkaline-earth-like and alkali atoms, and how they can be made internal state-
(or qubit-) dependent. We discuss these ideas on the single-particle level
initially, with interactions between atoms to be added in Sec.~\ref{sec4}.

As shown in Fig.~1a, our goal is to divide each period of a standard
$\lambda/2$ optical lattice into several potential wells with controllable
barrier heights and well depths in this subwavelength structure. A laser can
then be used to address atoms on different sites independently based on energy
selectivity, coupling to an unperturbed reference state. In addition, we can
control in this way the tunneling of atoms between adjacent wells in the
subwavelength structure.

The simplest way to achieve this goal is shown in Fig.~1b. We consider a
two-level atom with two long-lived internal states $\ket{e}$ and $\ket{g}$. We
assume that we can generate counter-oscillating \emph{off-resonant} potentials
for these two internal states, for which the implemenation will be discussed in
Sec.~\ref{sec3}. We then couple the internal states with additional dressing
fields, which generate avoided crossings in the background potentials, and thus
a subwavelength structure. Here, the well depth and barrier heights can be
controlled by varying the detunings and Rabi frequencies of the dressing
fields, which shifts the locations of the avoided crossings. In Fig.~1b, we
show the simplest case with isolated avoided crossings.

It is also possible to generate background potentials with \emph{near-resonant}
coupling between $\ket{g}$ and $\ket{e}$ with a Rabi frequency varying
sinusoidally in space (e.g., as generated by a standing wave light field). This
results in counter-oscillating potentials due to AC-Stark splitting, which can
then be coupled by a series of dressing fields to produce the desired
subwavelength structure (see Fig.~1c). This scheme is motivated, in particular,
by alkaline-earth-like atoms (as discussed in Sec.~\ref{sec3}).

\subsection{Addressable subwavelength lattices with two-level atoms in 1D: Off-resonant scheme\label{sec2a}}

We now discuss the details of the generation of subwavelength lattices with
off-resonant background potentials for a two-level atom. In the rotating wave
approximation with respect to the optical frequencies, we may write the single
atom Hamiltonian as:
\begin{eqnarray}
H&=&H_M+H_0+H_1,
\end{eqnarray}
with
\begin{eqnarray}
H_M&=&\frac{\hat{p}^2}{2m} \label{eq:hamiltonian}\\
H_0&=&V_g(x)|g\rangle\langle g|+(V_e(x)-\delta_1)|e\rangle\langle e|\nonumber\\
H_1&=&\sum_{n=1,2,...} \frac{\Omega_n}{2}\exp(-i(n-1)\delta t)|e\rangle\langle g|+{\rm h.c.}.\nonumber
\end{eqnarray}
Here, $H_M$ is the kinetic energy, $H_0$ describes the background potentials,
and $H_1$ contains the dressing fields with Rabi frequencies $\Omega_n$. For
dressing fields produced as an equally-spaced comb of sidebands, the detunings
can be written as $\delta_n=\delta_1+(n-1)\delta$, where $\delta_1$ is the
detuning of the first dressing field. In the off-resonant scheme, the
background potentials for the two internal states are generated by standing
wave fields and are opposite to one another. For this, we may take
$V_g(x)=-V_e(x)=-V_0\sin^2(kx)$, where $k=2\pi/\lambda$ is the wavenumber of
the laser generating the background potential.

Generation of the adiabatic potentials is based on the validity of a
Born-Oppenheimer-type assumption. This involves the fact that the kinetic
energy of the atoms is small on a scale given by the separation of the
resulting adiabatic potentials. The wavefunction $|\Phi(x,t)\rangle$ of a
single atom satisfies the Schr\"odinger equation,
\begin{equation}
i\hbar\frac{\partial}{\partial t}|\Phi(x,t)\rangle=(H_M+H_0+H_1)|\Phi(x,t)\rangle.
\end{equation}

If we omit the kinetic energy term from the Hamiltonian, we obtain an equation for adiabatic eigenstates, $\ket{\Psi_{\alpha}(x,t)}$,
\begin{equation}
i\hbar\frac{\partial}{\partial t}e^{-iE_{\alpha}(x)\,t}|\Psi_{\alpha}\rangle=(H_0+H_1)e^{-iE_{\alpha}(x)\,t}|\Psi_{\alpha}\rangle,
\end{equation}
where the $\ket{\Psi_{\alpha}(x,t)}$ are periodic functions and can be expanded
as a Fourier series. For a two-level atom, it takes the form
\begin{equation}
|\Psi_{\alpha}(x,t)\rangle=\sum_{l=-\infty}^{\infty} \exp(-il\delta t)[c_g^l(x)|g\rangle+c_e^l(x)|e\rangle].
\end{equation}
This results in a Floquet eigenvalue equation for $E_\alpha(x)$, which plays
the role of the adiabatic potentials,
\begin{eqnarray}
(V_g(x)-l\delta)c_g^l+\sum_n\frac{\Omega_n}{2} c_e^{l+n-1}=E_{\alpha}(x)c_g^l\\
(V_e(x)-\delta_1-l\delta)c_e^l+\sum_n\frac{\Omega_n}{2} c_g^{l-n+1}=E_{\alpha}(x)c_e^l.
\end{eqnarray}

The complete wavefunction can then be expanded in a basis of these adiabatic eigenstates, which play the role of Born-Oppenheimer channel functions,
\begin{equation}
|\Phi(x,t)\rangle=\sum_{\alpha}c_{\alpha}(x,t) |\Psi_{\alpha}(x,t)\rangle,
\end{equation}
resulting in the equation
\begin{equation}
i\hbar\frac{\partial}{\partial t}c_{\alpha}(x,t)=[H_M+E_{\alpha}(x)]c_{\alpha}(x,t)+\sum_{\beta}H_{M}^{\alpha\beta}c_{\beta}(x,t),
\end{equation}
where $H_M^{\alpha\beta}\equiv\langle\Psi_{\alpha}|H_M|\Psi_{\beta}\rangle$.
The equations for different $\alpha$ decouple provided that the mixing terms
coupling the channels are small. We expect this to be the case provided that
the kinetic energy of the atom in the lattice, which is typically on the order
of the recoil energy, is much smaller than the separation between the adiabatic
potentials, which is characterised by laser parameters, e.g. the Rabi frequency
and detuning. Below we will initially neglect these terms, and in
Sec.~\ref{sec5} we will discuss in detail non-adiabatic processes, and how they
should be suppressed in experimental implementations.

\begin{figure}[tbp]
\centerline{\includegraphics[width=8cm]{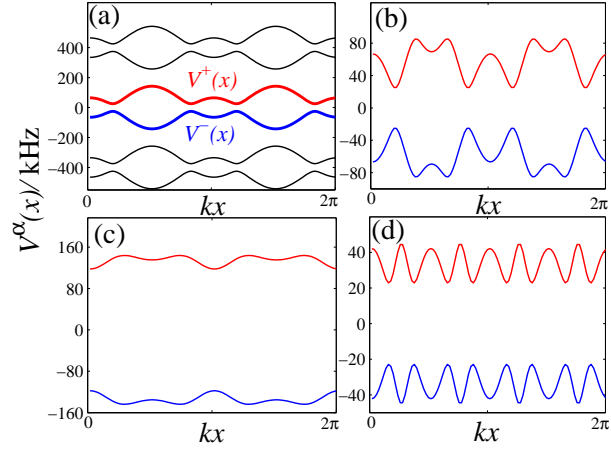}}
\caption[Fig. 2]{Illustration of the controllability of subwavelength potentials.
(a) Typical subwavelength lattice with $n=1$. The potential in the figure
is generated with the off-resonant scheme with the parameters:
$V_0=200$kHz, $\Omega_1=2\pi\times50$kHz, $\delta_1=120$kHz. Similar potentials can also be generated with
the near-resonant scheme. (b) Subwavelength double-well structure generated with the off-resonant
scheme with the parameters: $V_0=200$kHz, $\Omega_1=2\pi\times50$kHz, $\delta_1=120$kHz, $\Omega_2=2\pi\times50$kHz,
$\delta=220$kHz. The relative depth of the wells in the subwavelength structure can be tuned by changing $\Omega_2$ or $\delta$. (c)
Subwavelength double-well structure generated with the near-resonant scheme with the
parameters: $\Omega_0=2\pi\times400$kHz, $\delta_0=100$kHz, $\Omega_1=2\pi\times560$kHz, $\delta=-400$kHz.
The central barrier height in the lower dressed potential can be adjusted by changing
either $\delta$ or $\Omega_1$. (d) Subwavelength double-well structure generated with
the off-resonant scheme with $n=3$ dressing fields: $V_0=200$kHz, $\delta_1=60$kHz,
$\Omega_1=\Omega_2=\Omega_3=2\pi\times50$kHz, $\delta=140$kHz. }
\end{figure}

In the case of two-level atoms, the eigenvalues $E_\alpha(x)$ of the equations
appear in pairs, and may be identified as the upper $V^{+}(x)$ and the lower
$V^{-}(x)$ dressed potentials for a given Floquet manifold (see Fig.~2). The
corresponding eigenstates for the adiabatic potentials in general have
components from different Floquet manifolds. For the simple case of one
dressing field in the off-resonant scheme (with only $\Omega_1$ non-zero), the
different manifolds in the Floquet basis decouple, and we only need to
diagonalise a two-by-two matrix. In this case, the expression for the adiabatic
potential and its corresponding states are:
\begin{eqnarray}
V^{\pm}&=&-\frac{Z}{2}\pm\sqrt{X^2+Y^2}\nonumber \\
\left|\Psi^{\pm}\right\rangle&=&\frac{1}{2}[(1\mp\frac{X}{\sqrt{X^2+Y^2}})^{\frac{1}{2}}|g\rangle
\pm(1\pm\frac{X}{\sqrt{X^2+Y^2}})^{\frac{1}{2}}|e\rangle],\label{eq:dressedstates}
\end{eqnarray}
where $X\equiv V_e(x)-\delta_1/2$, $Y\equiv \Omega_1/2$, $Z\equiv \delta_1$.
The resultant dressed potential has a lattice spacing of $\lambda/4$. The
structure of the adiabatic potential becomes more complicated with more
dressing fields. With $n$ dressing fields, it is possible to create a lattice
with spacing approximately $\lambda/2(n+1)$. This increase in the spatial
resolution is at the expense of a decrease in the lattice depth.

\subsection{Addressable subwavelength lattices with two-level atoms in 1D: Near-resonant scheme\label{sec2b}}

In the near-resonant scheme, the counter-oscillating background potentials are
generated by directly coupling the two internal states with standing wave
fields, producing a pair of AC-Stark split states. The Hamiltonian describing
the coupling between states is now given by [cf. Eq.~(\ref{eq:hamiltonian})]
\begin{eqnarray}
H_0&=&-\delta_0|e\rangle\langle e|+\left(\frac{\Omega_0(x)}{2}|e\rangle\langle g|+{
\rm h.c.}\right)\nonumber\\
H_1&=&\sum_{n} \frac{\Omega_n}{2}\exp(-in\delta t)|e\rangle\langle g|+{\rm h.c.},
\end{eqnarray}
where $\Omega_0(x)=\Omega_0\sin(kx)$ is the Rabi frequency corresponding to a
standing wave field, and $\Omega_n$ ($n=1,2,...$) are the Rabi frequencies of
the dressing fields with corresponding detunings $\delta_n=\delta_0+n\delta$.
Note that the background potentials are associated with the eigenstates of
$H_0$, with the values of the the potentials and the states given by the same
form as Eq.~(\ref{eq:dressedstates}), with $X\equiv -\delta_0/2$, $Y\equiv
\Omega_0(x)/2$ and $Z\equiv\delta_0$ \cite{Kaza}.

Similar to the procedure in the previous section, we derive the equations for the adiabatic potentials,
\begin{eqnarray}
(-l\delta)c_g^l+\frac{\Omega_0(x)}{2}c_e^l+\sum_n\frac{\Omega_n}{2} c_e^{l+n}=E_\alpha (x)c_g^l\\
(-\delta_0-l\delta)c_e^l+\frac{\Omega_0(x)}{2}c_g^l+\sum_n\frac{\Omega_n}{2} c_g^{l-n}=E_\alpha (x)c_e^l,
\end{eqnarray}
with $E_\alpha(x)$ being the adiabatic potentials.

In Fig.~2 we give examples of the flexibility in constructing adiabatic
potentials via the near-resonant scheme as well as the off-resonant scheme,
including the realisation of  double-well structures on a subwavelength scale.
In Fig.~2b,c, we demonstrate the simple case of one dressing field: by varying
the Rabi-frequency or the detuning of the dressing field, the spatial structure
and the central barrier height of the subwavelength double-well structure can
be controlled. This is essential for entanglement manipulation and state
preparation, as we will discuss later.

\subsection{Adiabatic potentials in higher dimensions\label{sec2c}}
\begin{figure}[tbp]
\centerline{\includegraphics[width=8.5cm]{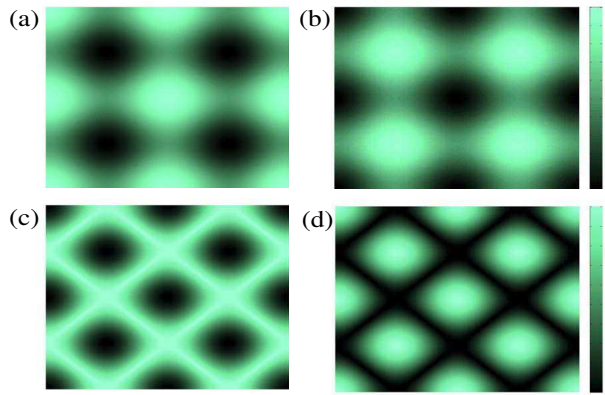}}
\caption{Adiabatic dressed potentials in 2D from an initial square lattice,
with depth $2V_0$, formed with a single dressing field, $\delta_1=2V_0$
(Note that the potential energy represented by each colour is particular to the figure part,
and in each case darker colours correspond to lower potential energy):
(a) The initial off-resonant potential for state $\ket{g}$, $V_g(x,y)=-V_0 \sin^2(k_l x)-V_0 \sin^2(k_l y)$.
(b) The initial resonant potential for state $\ket{e}$, $V_e(x,y)=-V_g(x,y)$.
(c) The lower dressed potential $V^-(x,y)$, corresponding to a square lattice
with $1/\sqrt{2}$ times the original period. (d) The upper dressed potential $V^+(x,y)$, corresponding to interconnected rings.
Note that the adiabatic potentials $V^\pm(x,y)$ can be found from Eq.~(10),
and have a maximum depth $V_0$, which decreases as the Rabi frequency of the dressing field increases.  }
\label{2ds}
\end{figure}

\begin{figure}[tbp]
\centerline{\includegraphics[width=8.5cm]{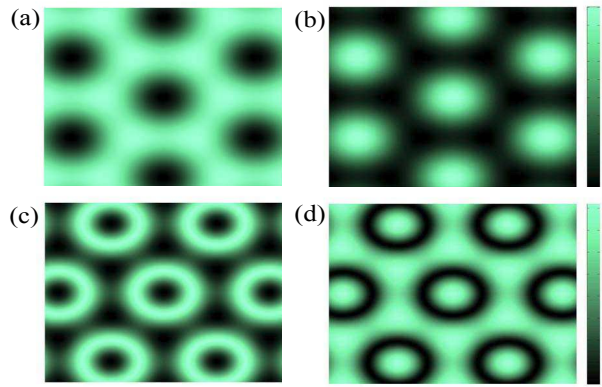}}
\caption{Adiabatic dressed potentials in 2D from an initial triangular lattice,
with depth $2V_0$, formed with a single dressing field, $\delta_1=2V_0$
(Note that the potential energy represented by each colour is particular to the figure part,
and in each case darker colours correspond to lower potential energy):
(a) The initial off-resonant potential for state $\ket{g}$, $V_g(x,y)$.
(b) The initial resonant potential for state $\ket{e}$, $V_e(x,y)=-V_g(x,y)$.
(c) The lower dressed potential $V^-(x,y)$, corresponding to a honeycomb
lattice with additional sites in the centre of a ring-shaped barrier.
(d) The upper dressed potential $V^+(x,y)$, corresponding to disconnected rings. Note that the adiabatic potentials $V^\pm(x,y)$
can be found from Eq.~(10), and have a maximum depth $V_0$, which decreases as the Rabi frequency of the dressing field increases.  }
\label{2dh}
\end{figure}

In higher dimensions, adiabatic potentials can also be used both to create
shorter period lattices, and to change the geometry of the lattices. Two
examples of 2D potentials are shown in Figs.~\ref{2ds}, \ref{2dh}. In
Fig.~\ref{2ds} we show the form of the dressed potentials obtained with an
initial square lattice with period $a$. From initial off-resonant potentials
$V_e(x,y)=V_0 \sin^2(k_l x)+V_0 \sin^2(k_l y)$, and $V_g(x,y)=-V_e(x,y)$, and
using a detuning $\delta_1=2V_0$, we obtain for $V^-(x,y)$ a square lattice of
period $a/\sqrt{2}$, as the minima in this dressed potential occur at each
position of a minimum in $V_e(x,y)$ or $V_g(x,y)$. In the upper adiabatic
potential $V^+(\mathbf{x})$, we observe the same form inverted, giving rise to
a potential consisting of interconnected rings.

In Fig.~\ref{2dh}, we show the dressed potentials arising from an initial
triangular  lattice, formed by three interfering laser beams at an angle of
$\pi/3$ to each other. This produces a well-defined Honeycomb lattice
structure, with auxiliary sites introduced in the lower dressed potential,
$V^-(x,y)$. In the upper dressed potential, $V^+(x,y)$ we observe a series of
disconnected rings (see Fig.~\ref{2dh}d). Such disconnected ring structures can
also be produced from a square 2D lattice with two dressing frequencies. Note
that the control over well depth here could have many useful applications,
e.g., in the case of the square lattice it could be well-suited to
implementation of ring-exchange Hamiltonians \cite{Buchler05}.

\section{State-dependent subwavelength potentials with multi-level atoms\label{sec3}}
The ideas for creating addressable, subwavelength lattices for two level atoms
as discussed in the previous section will now be extended to multi-level atoms,
where it is possible to create internal state- (qubit-) dependent potentials.
We will discuss, in particular, the case of alkaline-earth-like atoms with
nonzero nuclear spin, e.g. $^{171}$Yb. In addition to the optical excitation to
the metastable triplet levels, we have an extra degree of freedom in the form
of different magnetic sublevels, which arise from the nuclear spin.

\subsection{Off-Resonant Lattices\label{sec3a}}
\subsubsection{Lattice with Alkaline-earth-like atoms\label{sec3a1}}
As an example, we consider the level structure of $^{171}$Yb, for which the
nuclear spin $I=1/2$. We will make use of the states on the clock transition.
The states in the ground state manifold are $|{}^1S_0,m_I=\pm1/2\rangle$, and
those in the excited state manifold are $|{}^3P_0,m_F=\pm1/2\rangle$.  It is
also possible to generalise the scheme to other alkaline-earth-like atoms with
non-zero nuclear spin, e.g. $^{87}$Sr ($I=9/2$), though one then needs to
carefully choose the working states in the hyperfine structure
\cite{Boyd07,Boydthesis}.

\begin{figure}[tbp]
\includegraphics[width=16cm]{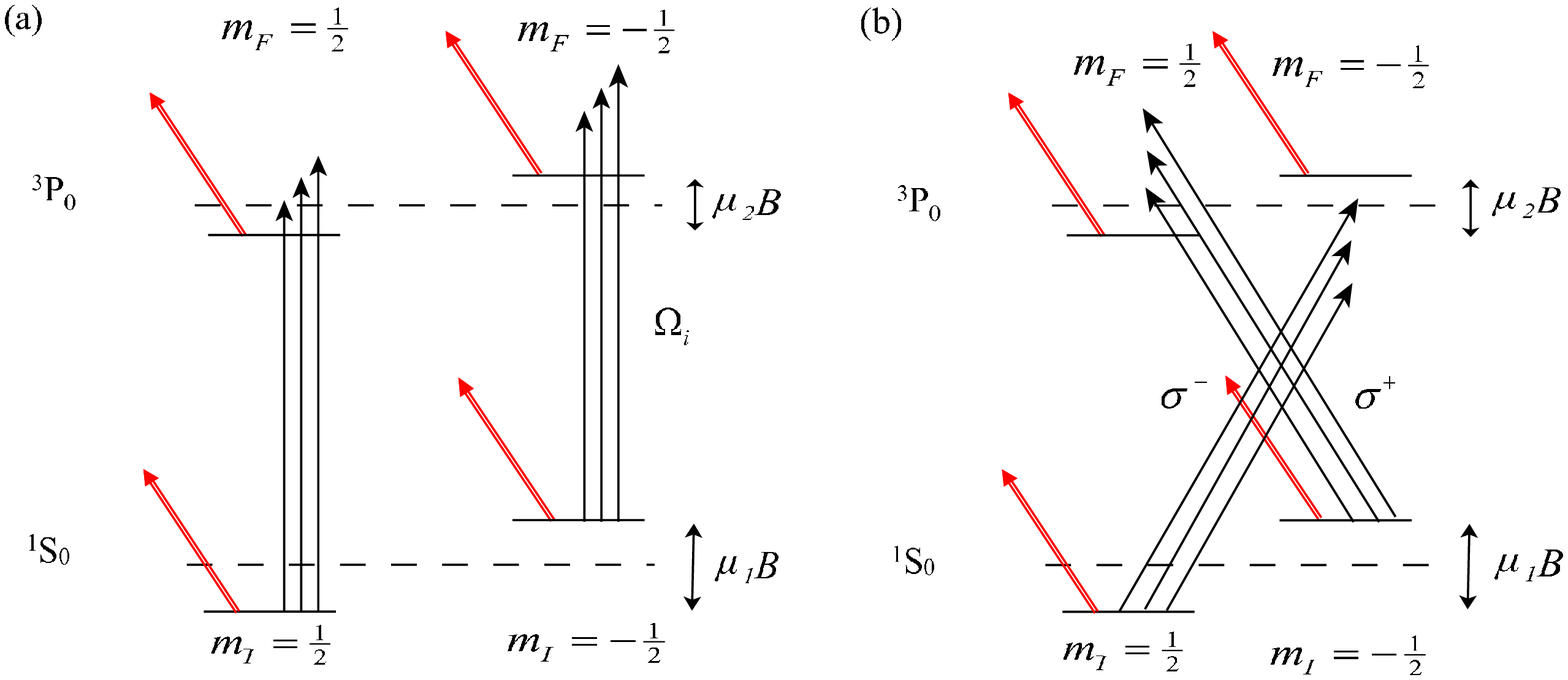}
\caption[Fig. 5]{Atomic level structure and laser configurations for
the generation of sub-wavelength lattices in $^{171}$Yb from off-resonant background potentials,
which are formed by AC-Stark shifts represented by the red arrows.
(a) Formation of subwavelength lattices in two two-level systems with $\pi$-polarised dressing fields (black arrows),
where the magnetic field induces differential Zeeman shifts in the $^1S_0$ ($\mu_1\sim
760$Hz$/$G), and the $^3P_0$ ($\mu_2\sim 350$Hz$/$G) manifolds
\cite{ytterbium0}. (b) Formation of subwavelength lattices in two two-level systems with circularly-polarised dressing fields (black arrows).
}
\end{figure}

To generate off-resonant subwavelength lattices, we induce opposite background
potentials for the states on the clock transition.  This is done by applying a
set of standing wave laser fields at a specific ``anti-magic'' wavelength at
which the AC polarisability of the manifolds $^1$S$_0$ and $^3$P$_0$ are
exactly opposite. This is exactly the opposite case as that used in optical
lattice clock experiments \cite{zeemanshift1,zeemanshift2,Ybclock}, where a
``magic'' wavelength is chosen so that the AC Stark shifts of the clock states
are equal. (We expect that the ``anti-magic'' wavelength in our case to be
$\sim 600$nm for $^{171}$Yb, and the resultant AC stark shift on the order of
several hundred kHz for a laser intensity of $I\sim 10$kW/cm$^2$.) Once the
background potentials are generated, we can couple these potentials by applying
additional optical dressing fields tuned near the resonance of the clock
transition $^1S_0-{}^3P_0$ at $\sim 578nm$. Atoms can be loaded into the minima
of the upper(lower) adiabatic potential by first preparing them in one of the
manifolds $^1$S$_0$ or $^3$P$_0$, and then appropriately ramping up the
dressing fields adiabatically.

As shown in Fig.~5a, the $^1$S$_0$ and $^3$P$_0$ manifolds for $^{171}$Yb are
each split into two sublevels. This gives us a four-level system, and dressing
fields coupling the different manifolds can be produced using either
$\pi$-polarised or circularly polarised light. In appendix A we summarise the
origins of the corresponding matrix elements \cite{ytterbium1,ytterbium2}.
Note, in particular, that the matrix elements for circularly polarised
couplings are non-zero as a result of contributions in the $^3$P$_0$ manifold
from other levels due to hyperfine mixing \cite{ytterbium1,ytterbium2}. In the
case that we use only circularly polarised dressing fields, we will produce two
independent two-level systems, where the resulting subwavelength lattice can be
manipulated for each by altering the appropriate component of each circular
polarization in the dressing field. For $\pi$-polarised light, we also obtain
two independent two-level systems, in which we can offset the relative well
depths by applying an external magnetic field. Due to the differential Zeeman
shift in the $^1$S$_0$ and $^3$P$_0$ manifolds \cite{ytterbium0}, this will
alter the detunings of the dressing fields, as well as the energies of the
uncoupled internal states. For example, with the laser parameters:
$V_0=200$kHz, $\Omega_1=2\pi\times200$kHz, $\delta_1=200$kHz, and under a small
magnetic field of $B=10$G, the estimated energy difference between the
potentials corresponding to the two different dressed states at $kx=\pi/2$ is
$\sim 7$kHz, and at $kx=\pi$ is $\sim 4$kHz. These numbers increase
significantly as the external magnetic field is increased. This allows the
state to be rotated between dressed states of the different two-level systems
selectively for different wells of the substructure using RF coupling fields.
This would be useful for applications to quantum information processing (see
below).

\begin{figure}[tbp]
\centerline{\includegraphics[width=8cm]{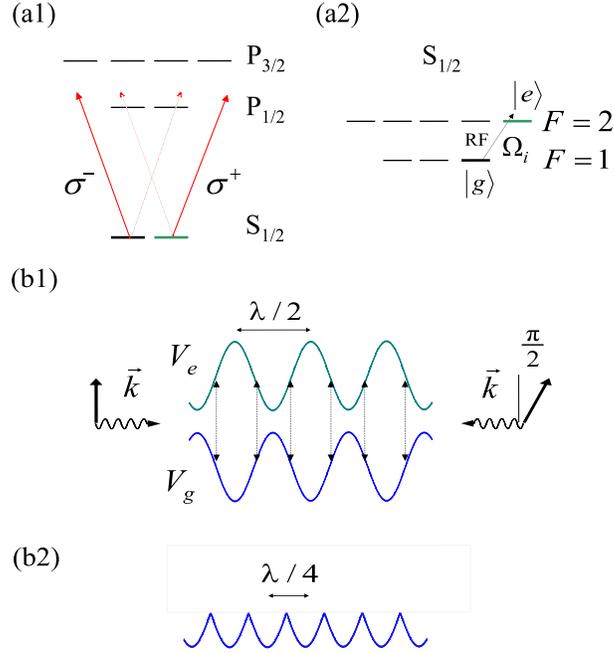}}
\caption[Fig. 6]{Schematics for the generation of an off-resonant
subwavelength lattice with $^{87}$Rb.
(a1) Lights of different circular polarisation are tuned near the centre of the fine structure splitting to
generate spin-dependent potentials. (a2)(b1) RF fields are applied to couple different hyperfine states with
counter-oscillating potentials, which are generated by shifting the spin-dependent potentials.
(b2) The resultant adiabatic potential in the case of one dressing field has a lattice spacing $\lambda/2$, with $\lambda$
the wavelength of the laser generating the spin-dependent potentials in (a1). }
\end{figure}

\subsubsection{Lattice with Alkali atoms\label{sec3a2}}

One may also generate off-resonant subwavelength lattices with alkali atoms.
Here, the counter-oscillating background potentials should be produced for two
different hyperfine ground states making use of well known ideas to produce
spin-dependent lattice potentials \cite{spinlattice1,spinlattice2}, as shown in
Fig.~6. These spin-dependent lattice potentials can then be coupled using RF
dressing fields to produce sub-wavelength optical lattices. Details of this
scheme are summarised in the caption of Fig.~6. Note that with alkali atoms it
is not straightforward to find combinations of states that act as independently
coupled two-level systems and that are simultaneously collisionally stable.
This will limit the application of alkali atoms in generating state-dependent
subwavenlength potentials, as well as in entanglement operations, as we will
discuss later for alkaline-earth-like atoms. In addition, with alkali atoms,
the two hyperfine ground states used to generate the dressed potentials
decohere fast in the presence of magnetic field fluctuations, which is not the
case with alkaline-earth-like atoms, where the dressed potentials are generated
almost purely from nuclear spins.

\begin{figure}[tbp]
\includegraphics[width=15cm]{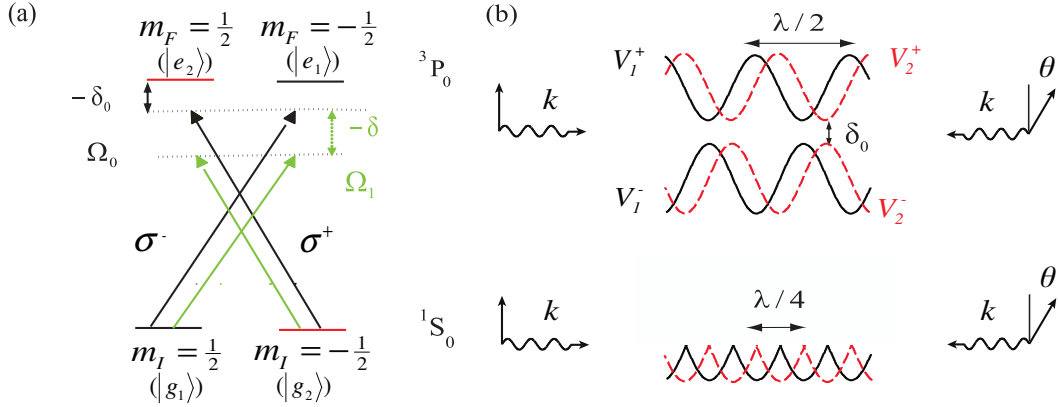}
\caption[Fig. 7]{(a) Level structure and laser configurations for the
generation of near-resonant subwavelength potentials with $^{171}$Yb
using cross-coupling dressing fields. (b) (above) With only the coupling
laser $\Omega_0$, there are two independent sets of spin-dependent lattice potentials $V^{\pm}_i$ ($i=1,2$),
where $i$ labels the two different sets of potentials created by lasers with different polarisation
and $\pm$ labels the upper/lower potential in a given set. The relative position of the two sets of potentials can be adjusted by varying $\theta$ in the lin-$\angle$-lin
configuration. The lattice spacing in this case is still $\lambda/2$.
(below) By adding one dressing field $\Omega_1$, the lower potentials become
state-dependent subwavelength potentials, with lattice spacing $\lambda/4$.
The relative positions of the two lower potentials can be adjusted by varying $\theta$.}
\end{figure}

\subsection{Near-Resonant Lattices\label{sec3b}}
\subsubsection{Lattice with Alkaline-earth-like atoms\label{sec3b1}}
The long-lived $^3$P$_0$ level motivates the application of near-resonant
lattices, as discussed previously in Sec.~\ref{sec2b}, with direct coupling with
a standing wave field tuned to the clock transition producing
counter-oscillating dressed potentials. The great advantage here, though, is
that we can now produce these background potentials with lights of two
different circular polarisations, generating two sets of independent background
potentials (see Fig.~7a). We can take advantage of this in two ways: (1) We can
couple these potentials with circularly polarised dressing fields (see Fig.~7),
so that we produce two independent sets of addressable sub-wavelength lattices;
or (2) We can couple these potentials with $\pi$-polarised fields, so that we
form dressed states with all four levels (see Fig.~8a), and use this to
engineer particular forms for the potentials. For the convenience of
discussion, we define the notation for the internal states
$|g_{1}\rangle=|{}^1S_0,m_I=+1/2\rangle$,
$|g_{2}\rangle=|{}^1S_0,m_I=-1/2\rangle$,
$|e_{1}\rangle=|{}^3P_0,m_F=-1/2\rangle$, and
$|e_{2}\rangle=|{}^3P_0,m_F=+1/2\rangle$.

In the first case, where the states are cross-coupled as shown in Fig.~7, we
then produce state-dependent lattices in the sense that we have two independent
two-level systems, in which we form dressed states of $\ket{g_i}$ and
$\ket{e_i}$($i=1,2$) for each $i$ independently. Both the near-resonant
background potentials and the final sub-wavelength potentials are
internal-state dependent, and can be easily manipulated independently. For
example, using lasers in a lin$\angle$lin configuration (two
counter-propagating, linearly polarized beams with a relative polarization
angle $\theta$), the lattices can be relatively shifted by varying $\theta$
\cite{spinlattice1,spinlattice2} (see Fig.~7b). Such spin dependence could be
used similarly to the case in Alkali atoms \cite{spinlattice1,spinlattice2},
and we expect losses due to spontaneous emissions to be strongly suppressed in
the present case. Note that the dipole matrix element for the cross coupling is
given by $2\pi\epsilon_0\hbar c^3\Gamma/(\omega_0^3)$, where $\omega_0$ is the
frequency of the clock transition $^1S_0-{}^3P_0$, and $\Gamma\sim 2\pi
\times10$mHz is the natural linewidth of the $^3P_0$ state in $^{171}$Yb
\cite{Ybclock,ytterbium0,ytterbium1,ytterbium2}(see appendix A for more
details).

\begin{figure*}[tbp]
\includegraphics[width=16cm]{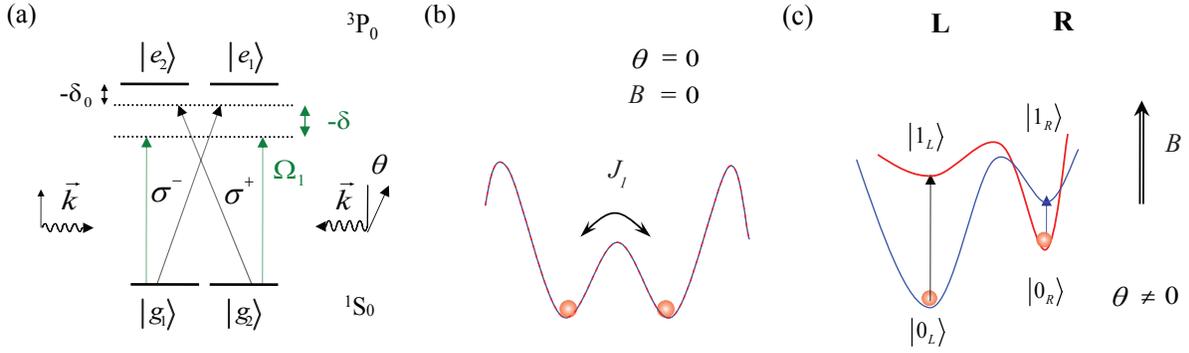}
\caption[Fig. 8]{(a) Level structure and laser configurations for the
generation of near-resonant subwavelength potentials with $^{171}$Yb using
$\pi$-polarised dressing fields. (b) Atoms in the subwavelength double-well
structure can be entangled by a controlled evolution with the exchange
interaction turned on. The interaction can be switched on and off by varying
the height of the central barrier. (c) In the presence of a small magnetic
field and as the angle in the lin-$\angle$-lin configuration $\theta\neq 0$,
atoms with different internal states in the double-well structure, which can be
defined as qubit states $|0_i\rangle,|1_i\rangle$ ($i={\rm L,R}$) as in the
figure, feel different potentials. Atoms within the subwavelength structure can
thus be selectively addressed.}
\end{figure*}

\begin{figure}[tbp]
\centerline{\includegraphics[width=8cm]{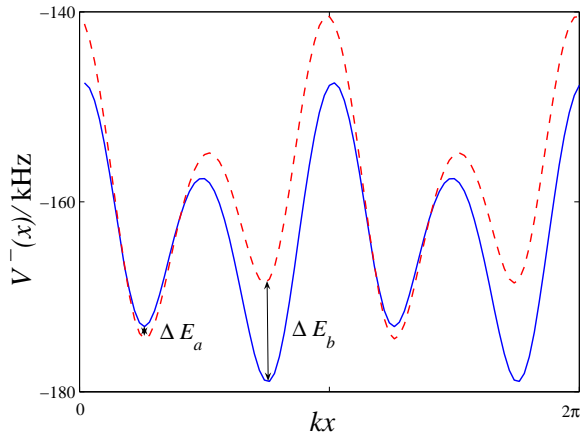}}
\caption[Fig. 9]{Biased double-well structure generated with the configuration in Fig.~8 for subwavelength addressing with $^{171}$Yb. The two
originally degenerate potentials (the dashed and the solid curve) are shifted relative to each other
under a small bias
magnetic field, and with a finite angle $\theta$ in the lin-$\angle$-lin configuration as shown in Fig.~8a.
Parameters: $\Omega_0\sim 2\pi\times400$kHz, $\delta_0\sim -100$kHz, $\delta\sim -420$kHz,
$\Omega_1=2\pi\times480$kHz, $\theta=0.1$, $B=10$G. The energy shifts are $\Delta E_a\sim1.3$kHz, and $\Delta E_b\sim 11$kHz,
respectively.}
\end{figure}
Subwavelength addressability can be achieved in this configuration in a manner
similar to that discussed in the previous section. By preparing the atoms near
the minima of the lower adiabatic potentials and applying an external magnetic
field to lift the degeneracy of the states in the subwavelength structure, we
can perform rotations between the independent two-level systems selectively by
well within the subwavelength structure.

In the second case we consider using $\pi$-polarised dressing fields as shown
in Fig.~8. Here, all the states are coupled to make the adiabatic potentials,
and changing the angle $\theta$ in the lin-$\angle$-lin configuration modifies
the shape of all four adiabatic potentials in each Floquet manifold. In this
configuration, it is possible to produce a repeated double-well potential, in
which the two wells can be individually addressed.  This is done by first
setting the angle $\theta$ in the lin-$\angle$-lin configuration to a small but
nonzero value. This will change the shape of the potentials and lift the
symmetry in the double-well structure. An example is given in Fig.~9, where at
$\Omega_0\sim 2\pi\times400$kHz, $\Omega_1\sim 2\pi\times480$kHz, $\delta_0\sim
-100$kHz, $\delta\sim -420$kHz, $\theta=0.1$, $B=10$G, the energy shifts are
$\sim 1.3$kHz and $\sim 11$kHz for the two sub-wells, respectively. Atoms in
the left and right hand side of the double-well structure can then be rotated
selectively by a RF coupling after the application of an external magnetic
field to shift the potentials.

In practice, strong coupling at intensity $I$ on the clock transition will also
give rise to AC-Stark shifts $\Delta E$ of the states $\ket{e_i}$ and
$\ket{g_i}$, which, as $\Delta E \propto I$ and $\Omega\propto \sqrt{I}$, will
become important at large fields (e.g, $\sim 10$kW/cm$^2$ for $^{171}$Yb
\cite{ybshift}). These have been neglected in the two-level atom picture
presented in Sec.~\ref{sec2b}. At higher fields, we essentially obtain a
combination of the off-resonant and near-resonant background potentials. If the
polarisability of $^3$P$_0$ and $^1$S$_0$ are opposite, it will simply enhance
the depth of the resulting potentials. These combined adiabatic potentials are
still spin-dependent as the AC-Stark shifts lose their spatial dependence as
the angle $\theta$ is increased to $\theta=\pi/2$. In the case that these
polarisabilities have the same sign, the AC Stark shift can also be cancelled
or modified by an additional laser at a different wavelength.

\subsubsection{Lattice with Alkali atoms\label{sec3b2}}

With alkali atoms one could also form near-resonant lattices by coupling
different hyperfine ground states in the $S_{\frac{1}{2}}$ level, e.g., via a
Raman process. If the Raman process beams have a sinusoidally varying effective
Rabi frequency, this will then generate two counter-oscillating dressed states.
These can be coupled by RF dressing fields to form subwavelength lattice
potentials. However, this does not give rise to simple internal state-dependent
lattices as in the case with alkaline-earth-like atoms.

\section{Application of subwavelength lattices in quantum simulation and quantum information\label{sec4}}

Having established the methods to realise subwavelength lattice potentials, we
now turn to discuss ways to utilise the flexibility in control of the
potentials for applications in quantum simulation and quantum information. This
requires consideration of the many-body physics in these potentials. We will
show how a generalised Hubbard model can be realised by engineering the
subwavelength potential. This generalised single-band Hubbard model features
fast tunnelling rates due to the small spatial separation of different wells in
the potential, which can be utilised to improve the timescales associated with
realisation of interesting quantum phases and entanglement via exchange
interactions.

\subsection{Single-band Hubbard model for the subwavelength lattice\label{sec4a}}

We begin by considering the simplest case of the off-resonant scheme with a
sinusoidal lattice in 1D, beginning with two background counter-oscillating
potentials $V_e(x)=-V_g(x)=V_0 \sin^2 (kx)$, as discussed in Sec.~\ref{sec2a}.
We then choose the detuning and Rabi frequency of  $n$ dressing transitions so
that the resulting bound states in each well of the adiabatic potential have
approximately the same energy. In this way a subwavelength lattice will be
formed in which the lattice spacing scales approximately as $\lambda/[2(n+1)]$,
while the depth $V_n$ of the potential can be as large as $V_n\sim V_0/(n+1)$
and decreases with increasing Rabi frequency $\Omega_n$. Motion can be
considered in the tight-binding limit of the resulting lattice model provided
$V_{n}/[(n+1)^2]\gtrsim 5E_R$, and the dynamics are described by the
single-band Hubbard model
\begin{equation}
\hat{H}=-\sum_{\langle\alpha,\beta\rangle}J_{\alpha\beta}c_{\alpha}^{\dagger}c_{\beta}
+\sum_{\alpha}\Delta_{\alpha}c_{\alpha}^{\dagger}c_{\alpha}
+\sum_{\alpha,\beta}U_{\alpha\beta}c_{\alpha}^{\dagger}c_{\beta}^{\dagger}c_{\beta}c_{\alpha}. \label{GenBH}
\end{equation}
Here, $c_{\alpha}$ is the annihilation operator for the bosonic or fermionic
atoms, and $\alpha$ specifying the lattice site and internal atomic (spin)
state, $J_{\alpha\beta}$ denotes the tunneling rates. We denote by
$\Delta_{\alpha}$ the energy offset for the internal state and lattice site
specified by $\alpha$, and $U_{\alpha\beta}$ are the collisional interaction
energies, which follow from our knowledge of the potentials and scattering
properties. A single-band Hubbard model is valid provided all of these energies
are smaller than the separation energy to higher bands, which, in the case that
the lowest bound states in all wells have the same energy, is bounded above by
the oscillation frequency in a typical well, $\omega=[4 d_1 V_0
E_R/(d_1^2+\Omega_1^2)^{1/2}]^{1/2}/h$.

If we choose the detunings of the $n$ dressing transitions so that
$\Delta_{\alpha}$ is constant, then in 1D, $J_{\alpha\beta}=J_n
\delta_{\alpha+1,\beta}$ can be estimated via a simple rescaling of the recoil
energy $E_R=\hbar^2k^2/2m$, as
\begin{equation}
\frac{J_n}{E_R} \sim 4\left(\frac{n+1}{\pi}\right)^{1/2}\left(\frac{V_{n}}{E_R}\right)^{\frac{3}{4}}
\exp\left(-\frac{2}{n+1}\sqrt{\frac{V_{n}}{E_R}}\right).
\end{equation}
Here, the tunneling rates increase exponentially with increasing $n$. Note,
however, that these rates can vary within the substructure, because the wells
with $n>1$ dressing frequencies will not be exactly symmetric, and this can
give rise to small splittings within the lowest Bloch band. In Fig.~10 we plot
values of $J_n$ for varying depths of the background potential, and number of
dressing transitions $n$. We will typically have onsite interactions only,
which can be controlled via the shape of each well and the scattering lengths
for atoms in the different internal states $\ket{e}$ and $\ket{g}$. This can
give rise to generalised Hubbard models, where the onsite interaction changes
values between neighbouring sites. Such models can support non-trivial phases,
such as coexisting Mott insulating and superfluid phases \cite{Jaksch98}. Note
that when there is only one dressing frequency $n=1$, the value of
$U_{\alpha\beta}$ is approximately the same as the corresponding onsite
interaction in the original background lattice potentials, as the curvature of
the wells is essentially unchanged.

\begin{figure}
\centerline{\includegraphics[width=8cm]{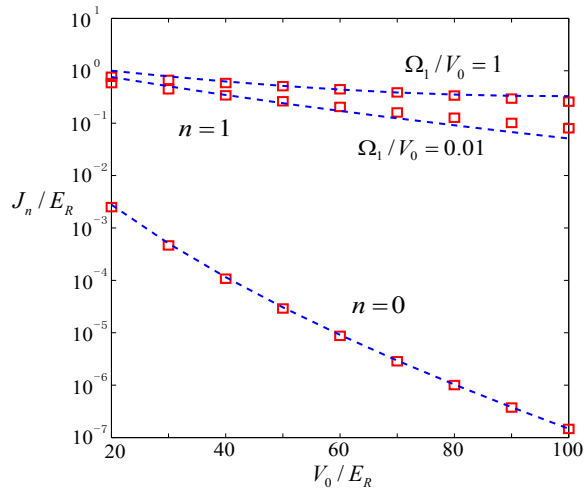}}
\caption[Fig. 10]{Tunneling rate $J_n$ in the presence of $n$ dressing fields as
a function of $V_0/E_R$. The squares are numerically-obtained values for the
case $n=0$(lowest curve) and $n=1$(upper two curves), respectively.
The dotted lines are plots of the $n$-dependent scaling function (see Eq.~(15) in the text).
The lowest curve is for the case $n=0$, and the higher two curves are for the $n=1$ case.}
\end{figure}

\subsection{Applications to Quantum Information\label{sec4b}}

As discussed in Sec.~\ref{sec3b}, we have with $^{171}$Yb the possibility to
form different addressable subwavelength lattices from four internal states. We
have different possibilities then to encode qubits and generate entanglement
depending on our choice of configurations for the dressing transitions.

In the case that we choose circularly polarised dressing transitions (see
Fig.~7), we obtain two independent two-level systems, for which we can control
the subwavelength lattices independently (see Sec.~\ref{sec3b}). We can then
encode qubits on the dressed states of the two two-level systems. These can be
addressed individually within the substructure of the subwavelength lattice
(i.e., modulo the period of the background potentials, where  other recent
ideas could be applied to provide addressing on the larger length scale
\cite{lukinaddress}). We can then make use of the state-dependent lattice
potentials to perform entangling operations, e.g., using controlled collisions
\cite{spinlattice1,spinlattice2}.

Another possibility arises in the case that we use $\pi$-polarised dressing
fields and form adiabatic eigenstates combining all four internal states (See
Sec.~\ref{sec3b} and Fig.~8). Here it is possible to create entanglement for
fermionic atoms such as $^{171}$Yb within the double-well structures using
exchange interactions \cite{exchangegate,Duan03}, where the fast tunneling
rates $J_{\alpha\beta}$ between neighbouring wells give rise to faster gate
times than are possible without subwavelength lattices. This leads to quantum
gate schemes in the spirit of the Loss-DiVincenzo proposal for quantum dots
\cite{Burkard}, which is based on encoding qubits on electron spins in
electrically gated quantum dots, and performing entangling operations using
exchange interactions between neighbouring dots by ramping down the
corresponding potentials. The same techniques can be applied here. We consider
the two dressed states of our four initial internal levels that have the same
dressed potentials, and define these to be our qubit states. We then consider
the dynamics generated by the Hamiltonian in Eq.~(\ref{GenBH}), but now with
the tunnelling parameters $J_{\alpha\beta}$ determined by our choice of
subwavelength lattices with repeated double-well potentials. The tunnelling
between these wells can be made negligible on the timescale of the experiment,
so that the dynamics within each double well is independent. We then have a
situation very similar to that discussed in the analysis of spin-exchange
gates, e.g., in Ref.~\cite{exchangegate}, leading from the two-site double-well
potential to exchange interactions between the spin states in each well, given
by $H_{\rm SE}= J_1 \vec \sigma_L. \vec{\sigma}_R $, where $J_1\equiv
J^2(1/U_L+1/U_R)$. Here, $\sigma_L$ and $\sigma_R$ represent the spin operators
in the left and right well respectively, where the spin in each well
corresponds to two-state system consisting of the two different dressed states
we are condsidering. The coefficient $J_1$ is determined from the tunneling
amplitude between sites, $J=J_{\alpha\beta}$ [see Eq.~(\ref{GenBH})]; and the
onsite interaction strengths $U_L$ and $U_R$, between two atoms, one in each
state, in the left well and right well respectively.

In the context of electrically gated quantum dots, one of the most challenging
aspects of implementing quantum computing is to achieve local addressing of the
qubits. As described above, this is made possible here by the addressability of
the individual wells within the subwavelength structure, as discussed in
Sec.~\ref{sec3b}. There we specifically discussed rotation between different
qubit states. However, it is also possible to perform readout state-selectively
by creating an energy offset between the two qubit states by introducing an
external magnetic field, and then coupling them state-selectively to the
$^3$P$_2$ level. This is possible because the $^3$P$_2$ level has a very small
natural linewidth $\Gamma\sim 2\pi \times15$mHz \cite{ytterbium1}, so that
coupling to this level can be easily made frequency-selective. Coupling between
the qubit states and $^3$P$_2$ can be generated with a Raman pulse, making use
of the larger coupling elements of both $^3$P$_2$ and $^3$P$_0$ to high-lying
states such as $^3$S$_1$.

\section{Landau-Zener-type Losses\label{sec5}}

As discussed in Sec.~\ref{sec2a}, one of the major limitations of the adiabatic
potentials is the loss of atoms via Landau-Zener-type processes. These losses
correspond to the tunneling of atoms from a given adiabatic potential into the
continua of other adiabatic potentials. In the following we show that these
processes can be made exponentially small by increasing the laser power.

Landau-Zener-type losses can be understood by looking at the simple scenario of
a single avoided crossing of two adiabatic potentials. This problem has been
discussed in detail in Ref.~\cite{Kaza}, where the two adiabatic potentials
$\pm U(x)$ are generated by coupling a two-level atomic system using a standing
wave light field with spatially oscillating Rabi-frequency. The loss rate
$\Gamma$ of a particle from the bottom of a well of the upper adiabatic
potential $U(x)$ into the continuum spectrum of $-U(x)$ can be estimated
semiclassically for a small violation of adiabaticity:
\begin{equation}
\Gamma\sim \Gamma_0 \exp\left[2 i\int_0^{x_0} (p_{+}-p_{-})\right],\label{lossrate}
\end{equation}
where $p_{\pm}=\sqrt{2m[E\mp U(x)]}$, $E$ is the energy of the atom, and $x_0$
is determined by the saddle-point condition $p_{+}(x_0)=p_{-}(x_0)$. The
quantity $\Gamma_0$ is the attempt frequency, which is of the order of the
oscillation frequency $\omega$ of a particle at the bottom of the upper
potential $U(x)$ \cite{Coleman1,Coleman2}. This analysis, in particular
Eq.~(\ref{lossrate}), can be directly adopted for the estimation of the
Landau-Zener-type loss rate from the upper adiabatic potential in the simple
case of one dressing field in the off-resonant scheme, and no dressing field in
the near-resonant scheme.

For the off-resonant scheme with only one dressing field, and for a detuning
$\delta_1=V_0$, the adiabatic potentials can be written as $\pm U(x)$, where
$U(x)=\sqrt{V(x)^2+\Omega_1^2/4} $, and $V(x)=V_0\sin(2kx)/2$. In the vicinity
of the saddle point $x_0=0$, the background potential can be linearised
$V(x)\sim V_0kx$, and we take the energy of the atom $E=U(x_0)$. We may then
Taylor expand the adiabatic potentials near the saddle point, $U(x)\sim
\Omega_1/2+m\omega^2x^2/2$, where $\omega\equiv2\sqrt{V_0^2E_R/\Omega_1}$ with
$E_R=\hbar^2k^2/2m$ being the recoil energy in the original normal lattice.
With these, one may carry out the integral in Eq.~(\ref{lossrate}) explicitly
and we get for the loss rate:
\begin{equation}
\Gamma\sim \Gamma_0\exp\left[-\alpha\frac{\Omega_1}{\omega}\right]=
\Gamma_0\exp\left[-\alpha\frac{\Omega_1^{3/2}}{2V_0E_R^{1/2}}\right],
\end{equation}
where the dimensionless parameter $\alpha\simeq 1.32$ in this case. Note that
$\Omega_1$ characterises in this case the energy separation between the
adiabatic potentials. Eq.~(17) is different from the scaling relation of a
typical Landau-Zener transition, where the exponent in the loss rate scales
with the energy separation squared. One may recover this typical scaling
relation by examining the loss rate of atoms with higher energies in the
adiabatic potential.

For the near-resonant lattices, different manifolds in the Floquet basis only
decouple in the absence of dressing fields. In this case, we have only the
background potentials $\pm U(x)$ resulting from the AC Stark splitting, where
$U(x)=\sqrt{\delta_0^2/4+\Omega_0(x)^2/4}$. We may also integrate
Eq.~(\ref{lossrate}) to calculate the loss rate from the upper background
potential:
\begin{equation}
\Gamma\sim\Gamma_0\exp\left[-\alpha\frac{\delta_0}{\omega'}\right]=
\Gamma_0\exp\left[-\alpha\frac{\delta_0^{3/2}}{\Omega_0E_R^{1/2}}\right]
\end{equation}
where $\omega'\equiv \sqrt{\Omega_0^2E_R/\delta_0}$, with $\alpha\simeq 1.32$.
In this case, $\delta_0$ characterises the energy separation between the
background potentials.

These results suggest that the loss rate from the upper adiabatic potential can
be made exponentially small by increasing the ratio of the energy difference of
the two potentials and the frequency of the atomic motion in the trap. In the
simple cases above, this translates to large Rabi-frequency $\Omega_1\gg E_R$
in the off-resonant case, and large detuning $\delta_0\gg E_R$ in the
near-resonant case.

\begin{figure}[tbp]
\includegraphics[width=8cm]{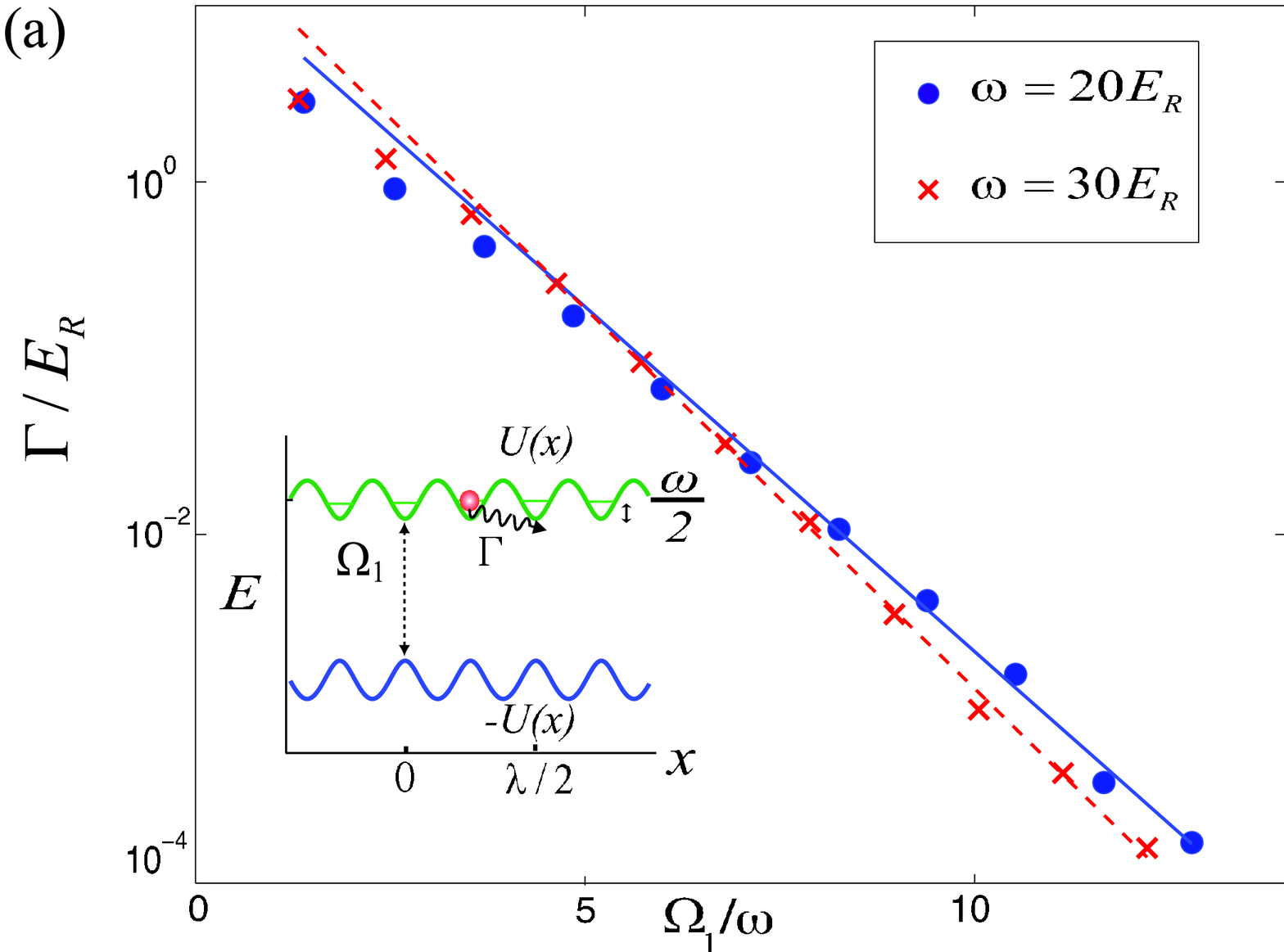}
\includegraphics[width=8cm]{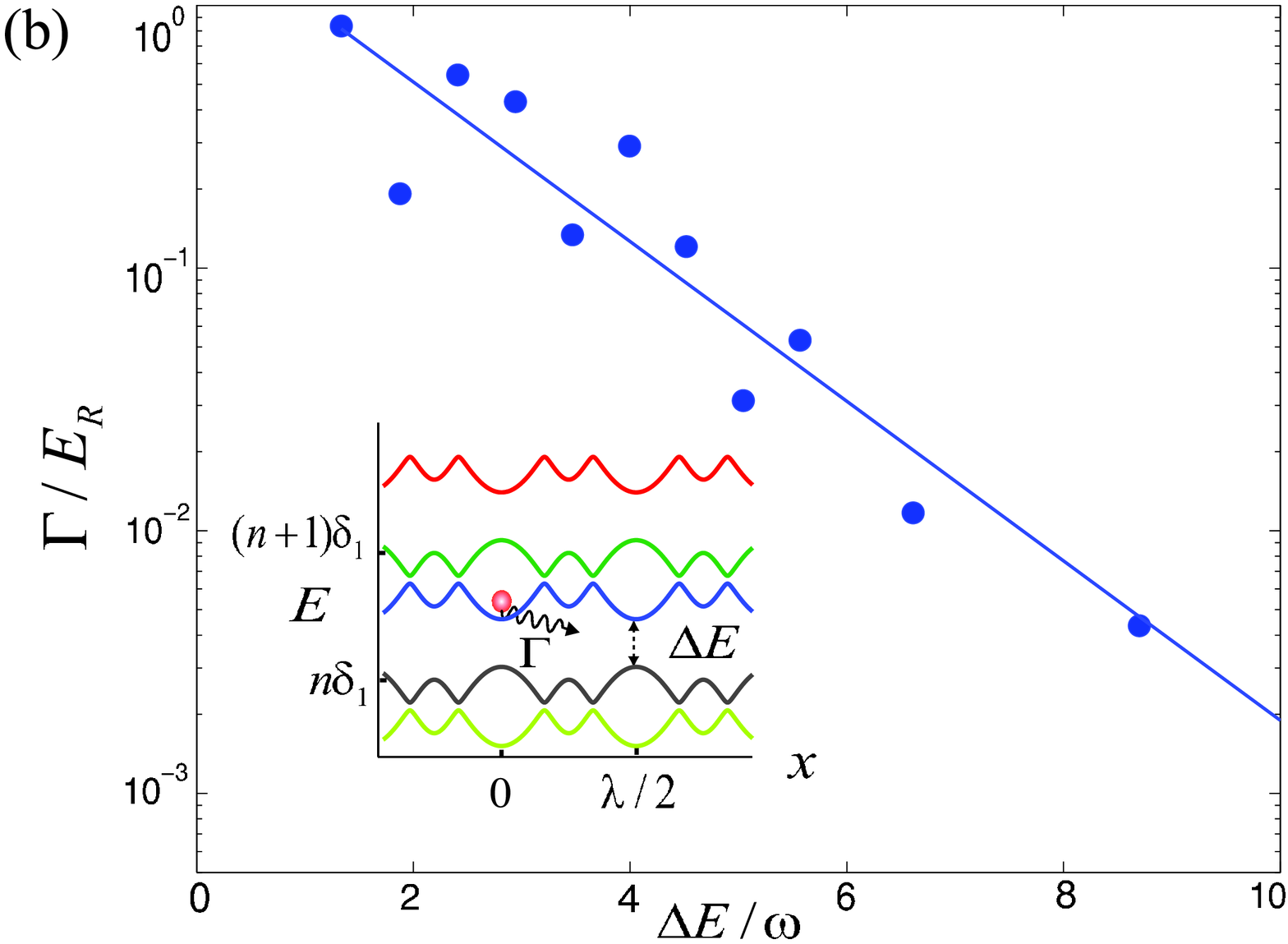}
\caption[Fig. 11]{Landau-Zener Losses from a $\lambda/4$ lattice, formed with one dressing transition: (a)  Numerical simulation results
for loss rates for atoms in the bound state of the
higher-energy adiabatic potential in the off-resonant scheme. The solid line is an exponential fit to the result. (inset) The corresponding adiabatic
potentials with parameters: $V_0=35E_R$, $\Omega_1= 30 E_R$, $\delta_1+\delta=35E_R$. (b) Numerical calculations of the decay rate
for atoms in the narrower wells in the low energy adiabatic potential in the near-resonant scheme. (inset) Typical potentials in the limit
where the Floquet manifolds are weakly coupled ($\Omega_1 \ll
\Omega_0$), with a series of Floquet manifolds each containing
2 dressed potentials. Parameters: $\Omega_0=31E_R$, $\delta_0=-15E_R$,
$\Omega_1=2E_R$, and $\delta_1=-\delta_0+(\sqrt{\delta_0^2+\Omega_0^2}+\delta_0)\times
0.8$. }
\end{figure}

When more complicated lattice potentials are considered, the loss rate due to
this Landau-Zener-type transition should in general be computed numerically.
However, simple estimates of the loss rates can be obtained by assuming that
for a given adiabatic potential the loss is dominated by the leak into the
continua of the adiabatic potentials which are closest in energy. If we assume
that the loss into the continuum of each potential is independent, the matrix
elements for the decay into the continuum of any one of these nearest adiabatic
potentials will have the form Eq.~(\ref{lossrate}) under the semiclassical
approximation. An exponential dependence of the decay rate $\Gamma$ on the
energy separation between the adiabatic potentials is then obtained, similar to
Eqs.~(17), (18). Numerically, we find the dimensionless parameter $\alpha$ to
be on the order of unity or larger when the laser parameters, e.g.
Rabi-frequencies and detunings, are much larger than the recoil energy $E_R$.
This implies that the loss rate $\Gamma$ can always be made exponentially small
by increasing the laser power.

To check the general conclusion drawn above, we have also performed numerical
simulations of the real time evolution of the atoms trapped in the adiabatic
potentials by integrating the Schr\"odinger equation for the full Hamiltonian
in the Floquet basis. The atom is initially prepared in the ground state of a
potential well, and transition rates are extracted from the projection of the
wavefunction onto the manifold corresponding to the initial adiabatic potential
as a function of time. For the simple case in Fig.~11a, we vary $\Omega_1$ for
fixed trapping frequency in the wells of the adiabatic potential
$\omega=2\sqrt{V_0^2 E_R/ \Omega_1}$. For a large gap between the potentials,
the loss rate scale exponentially as $\Gamma/E_R \propto \exp(-\alpha
\Omega_1/\omega)$, as shown by the exponential fit (solid line), which is in
agreement with the semiclassical calculation. From the fit, the factor $\alpha$
is estimated to be $\alpha\sim 0.9$ and $1.2$ for the exact and linearised
potentials, respectively, slightly smaller than the results in the
semiclassical approximation.  In the case of near-resonant scheme, as in
Fig.~11b, we fix $\sqrt{\Omega_0^2 E_R/ |\delta_0|}=30E_R$, and vary the energy
gap $\Delta E$ between the adiabatic potentials. We again observe an
exponential scaling, $\Gamma\propto \exp(-\alpha' \Delta E/ \omega)$ (solid
line), where the trapping frequency $\omega \approx 15 E_R$. In conclusion, we
see one can always avoid large Landau-Zener-type losses from the adiabatic
potential by increasing the laser power to make the background potential
sufficiently deep.

\section{Summary and Conclusion\label{sec6}}
To summarise, we have discussed various schemes for the generation of
subwavelength lattices with either alkali or alkaline-earth-like atoms. The
lattice spacing in these adiabatic potentials can be much smaller than the
wavelength of the laser that creates the lattice potential, depending on the
number of the additional dressing lasers. The enhanced tunneling rate one gets
from the reduced lattice spacing may be useful for the engineering of
generalised Hubbard models for quantum simulation, or for the design of fast
quantum gates for quantum computation. The adiabatic potentials in dimensions
higher in 1D may also have interesting structures, i.e. disconnected ring
potentials in 2D, which is useful for the realisation of interesting phases.
The first experimental steps in this direction have been recently been reported
in Ref.~\cite{PortoNew}. While we have discussed addressable lattices within a
$\lambda/2$ period, addressing on a larger scale can be introduced using
magnetic or electric field gradients, or other recently proposed techniques
\cite{sarma06,lukinaddress,Cho07,Rey07}. This, combined with the gate
operation, holds the prospect for the realisation of general quantum
computation in this system. We have also discussed in detail the
Landau-Zener-type losses from the adiabatic potentials. We have shown that the
loss rate can always be made small at the expense of laser power. For large
number of dressing fields, i.e. for small lattice spacing, the
Landau-Zener-type loss would become considerable, and would be the practical
limitation on how small a lattice spacing one can get in generating the
subwavelength lattices.

We thank H.~P.~B\"uchler, A.~V.~Gorshkov, S.~Kotochigova, I.~Lesanovsky,
M.~D.~Lukin, A.~Micheli, V. ~Pal'chikov, T.~Porto, Y.~Takahashi, J.~Ye, and
T.~Zelevinsky for interesting discussions. This work was supported by the
Austrian FWF through project SFB F15, and by the EU networks OLAQUI and SCALA.

\appendix
\section{Alkaline-Earth-Like Atoms and the Clock Transition}

In this appendix, we examine in detail the matrix element for the clock
transition $^1S_0-{}^3P_0$ in alkaline-earth-like atoms. For a detailed
discussion, please refer to Ref. \cite{ytterbium1,ytterbium2}.

We consider the matrix element $\langle ^1S_0,Im_I|D|^3P_0,F m_F\rangle$, where
$D$ is the electric-dipole operator. In the presence of a hyperfine
interaction, $|^3P_0,F m_F\rangle$ is not an eigenstate of the system. Under
the first order perturbation, the correction to the hyperfine sublevels in
$^3P_0$ manifold can be written as:
\begin{equation}
\sum_{\gamma}|\gamma,J,F m_F\rangle
\frac{\langle \gamma,J,F m_F|V_{\rm hf}|^3P_0,F
m_F\rangle}{E(\gamma)-E(^3P_0)},
\end{equation}
where $\gamma$ labels the manifolds that the the state is coupled to, and
$V_{\rm hf}$ is the hyperfine interaction. Due to the scalar nature of $V_{\rm
hf}$, the total angular momentum $F$ and its projection on z-axis $m_F$ are
good quantum numbers, as is shown in the expression above. As we are only
interested in the E1 transitions from $^1S_0$ to $^3P_0$, we only consider the
nuclear magnetic dipole contribution in the hyperfine interaction, which fixes
$J=1$ in the expression above.

The matrix element for the clock states then becomes:
\begin{eqnarray}
&&\langle ^1S_0,Im_I|D|^3P_0,Fm_F\rangle\nonumber\\
&&=\sum_{\gamma}\langle ^1S_0,Im_I|D|\gamma,J=1,Fm_F\rangle A\nonumber\\
&&=\sum_{m_I+m_J=m_F}^{\gamma}C^{F}_{m_Jm_I}\langle ^1S_0Im_I|D|\gamma,m_Im_J\rangle A,
\end{eqnarray}
where $A\equiv \frac{\langle \gamma,J,F m_F|V_{hf}|^3P_0,F
m_F\rangle}{E(\gamma)-E(^3P_0)}$, and $C^F_{m_Jm_I}$ is the Clebsch-Gordon
coefficient for the state $|\gamma,J=1,F\rangle$.

For the convenience of discussion, we now consider the concrete example of
$^{171}$Yb, which has a nuclear spin of $I=1/2$. It is obvious that the matrix
element is nonzero for both the $\pi$-polarised laser field and the
circularly-polarised laser field. For $\pi$-polarised lasers, the matrix
element is effectively given by
\begin{equation}
\langle\gamma,J=1,m_J=0,m_I=\pm\frac{1}{2}|D|^1S_0, m_J=0,m_I=\pm \frac{1}{2}\rangle;\nonumber
\end{equation}
while for circularly-polarized lasers, the matrix element is effectively given
by
\begin{equation}
\langle \gamma,
J=1,m_J=\pm1,m_I=\mp\frac{1}{2}|D|^1S_0,m_J=0,m_I=\mp\frac{1}{2}\rangle,\nonumber
\end{equation}
for the $\sigma^+$- and $\sigma^-$- polarised light, respectively.

The ratio between the matrix elements in the two cases is given by that of the
corresponding Clebsch-Gordon coefficients in Eq.~(A2).  One may further relate
these matrix element to the natural linewidth $\Gamma$ of the clock transition:
\begin{equation}
|\langle D\rangle|^2=\frac{3\pi\epsilon_0\hbar c^3}{\omega_0^3}\beta\Gamma,
\end{equation}
where $\omega_0$ is the frequency for the clock transition. The constant factor
$\beta$ comes from the Clebsch-Gordon coefficient, which has $\beta=1/3$ for
$\pi$-polarised lasers and $\beta=2/3$ for circularly polarised lasers. For
$^{171}$Yb, $\Gamma\sim2\pi\times 10$mHz \cite{Ybclock,ytterbium0,ytterbium1},
and a typical laser intensity of $I=10$kW/cm$^2$ gives a Rabi-frequency of
$\sim 2\pi\times220$kHz for a circularly polarised field.

\end{document}